\documentclass[runningheads]{llncs}
\ifdefined\pdfpagewidth
\fi
\usepackage[T1]{fontenc}
\usepackage{graphicx}
\usepackage{url}
\usepackage{amsmath}
\usepackage{amssymb}
\usepackage{booktabs}
\usepackage[table]{xcolor}
\usepackage{multirow}
\usepackage{bm}
\usepackage{placeins}
\usepackage{hyperref}
\usepackage{bbding}
\hypersetup{hidelinks}

\usepackage{enumitem}
\setlist[itemize]{topsep=2pt,partopsep=0pt,itemsep=0pt,parsep=0pt}
\newcommand{\methodname}{\mbox{MOTIF}}
\definecolor{GroupGray}{gray}{0.86}
\newcommand{\grouprow}[1]{\rowcolor{GroupGray}\multicolumn{7}{@{}l}{\textit{#1}} \\}

\begin{document}

\title{MOTIF: Motivation-guided Topology Inference for Cold-start Multimodal Recommendation}
\titlerunning{MOTIF for Cold-start Multimodal Recommendation}
\author{Yurui Shi\inst{1}\textsuperscript{\ensuremath{\dagger}} \and
Yuchen Miao\inst{2}\textsuperscript{\ensuremath{\dagger}}\textsuperscript{\Envelope} \and
Ximing Hu\inst{1} \and
Zijun Wang\inst{2} \and
Chang Han\inst{2}}
\authorrunning{Y. Shi et al.}

\institute{Taiyuan University of Technology, Taiyuan, China\\
\and
Sydney Smart Technology College, Northeastern University, China\\
\textsuperscript{\ensuremath{\dagger}}Co-first authors; \textsuperscript{\Envelope}Corresponding author.}

\maketitle
\thispagestyle{headings}

\begin{abstract}
Cold-start multimodal recommendation faces three coupled challenges: (i) sparse interactions obscure user intent, (ii) cold items remain topologically isolated, and (iii) similarity-based item graphs may cause semantic drift. To address these issues, we propose MOTIF, a Motivation-guided Topology Inference framework for cold-start multimodal recommendation. MOTIF integrates Semantic Motivation Reasoning, Knowledge-enhanced Graph Reconstruction, Weighted Graph Contrastive Learning, and Semantic-Structural Alignment. It uses offline LLM reasoning to infer motivation semantics, reconstructs transferable item-item topology, and learns robust graph embeddings without injecting generated text into prediction. Experiments on three multimodal benchmarks show consistent gains over graph-based, multimodal, cold-start, and LLM-enhanced baselines, with up to 6.07\% relative improvement over the strongest recent baseline.
\end{abstract}

%%%%%%%%%%%%%%%%%%%%%%%%%%%%%%%%%%%%%%%%%%%%%%%%%%%%%%%%%%%%

\section{Introduction}

Multimodal recommender systems use item content such as text, images, videos, and attributes to enrich preference modeling. However, cold-start recommendation remains difficult: cold users provide limited behavioral evidence, and cold items have insufficient interaction edges for graph propagation. Effective cold-start multimodal recommendation therefore requires both semantic understanding of sparse evidence and topological reconstruction of incomplete structures.

Existing methods address this problem through graph collaborative filtering~\cite{he2020lightgcn}, graph contrastive learning~\cite{wu2021sgl,lin2022ncl,yu2022simgcl}, multimodal representation learning~\cite{wei2019mmgcn,zhang2021lattice,zhou2023freedom}, cold-start adaptation~\cite{volkovs2017dropoutnet,lee2019melu,lu2020metahin,zhao2022cvar}, and LLM-enhanced semantic modeling~\cite{geng2022p5,bao2023tallrec,wei2024llmrec,ren2024rlmrec,di2026lmmrec}. Yet semantic modeling, graph reconstruction, and representation robustness are often handled separately, which limits their ability to jointly address sparse user intent, cold-item isolation, and noisy multimodal semantics.

\begin{figure*}[t]
\centering
\includegraphics[width=0.96\textwidth]{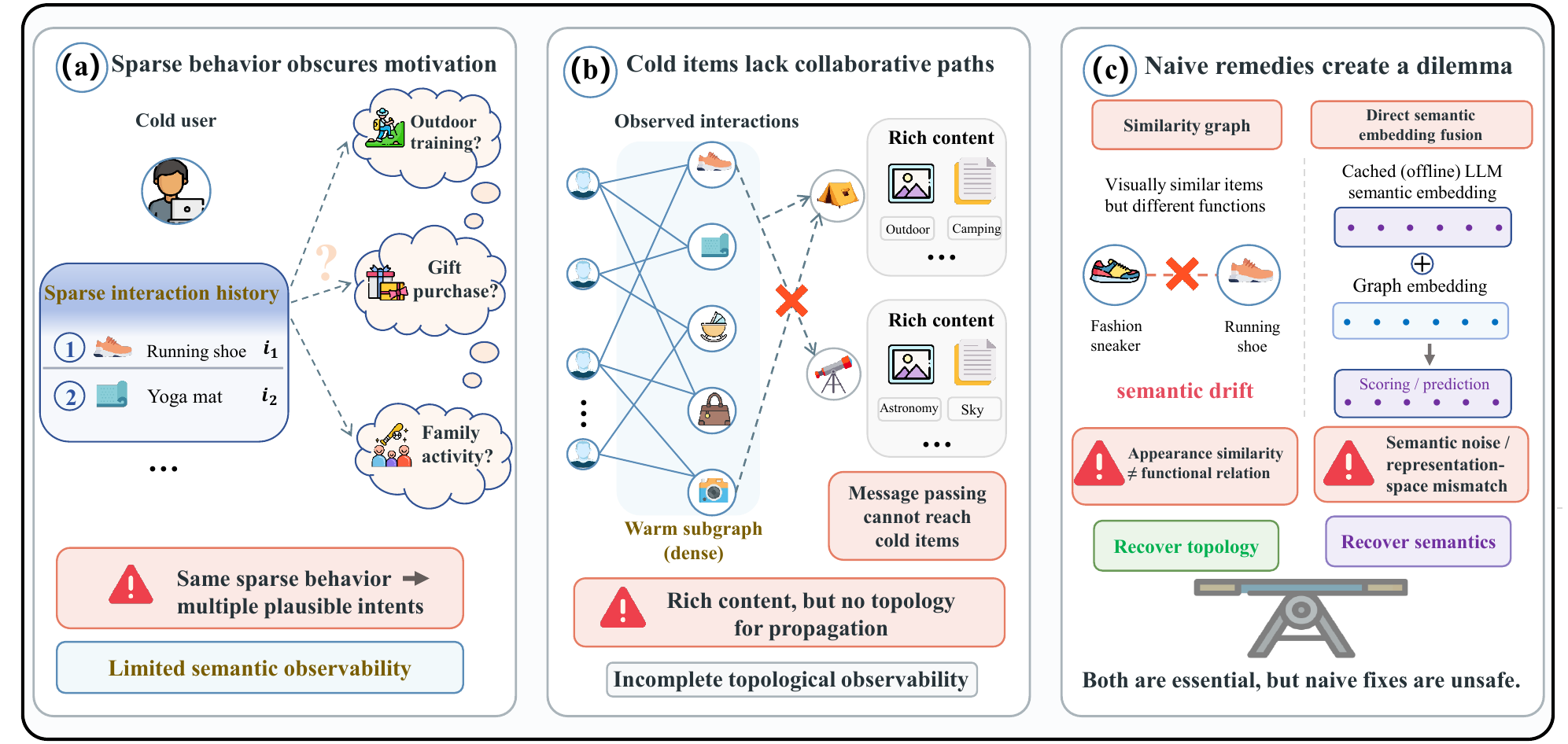}
\caption{Motivation of semantic-topological co-reconstruction for cold-start multimodal recommendation.}
\label{fig_intro_motivation}
\end{figure*}

Figure~\ref{fig_intro_motivation} illustrates the motivation. Cold-start multimodal methods face limited semantic observability and incomplete topological observability. Sparse interactions may not reveal user motivations, while similarity-based item graphs may connect visually similar but functionally different items, causing semantic drift. Directly feeding LLM-generated text into prediction can also introduce semantic noise. These observations motivate a framework that infers latent motivations, converts them into transferable graph connectivity, and learns robust representations on the reconstructed topology.

To this end, we propose MOTIF, an LLM-enhanced semantic-topological co-reconstruction framework for cold-start multimodal recommendation. MOTIF infers user and item motivations from sparse multimodal contexts, transforms them into knowledge-enhanced item-item topology for cold-item propagation, and learns robust representations through weighted graph contrastive learning and semantic-structural coordination.

The main contributions of this work are summarized as follows:
\begin{itemize}
\item We formulate cold-start multimodal recommendation as joint semantic and topological reconstruction over motivations and graph connectivity.
\item We use LLM-based motivation reasoning to build cold-start-aware item-item topology for collaborative propagation.
\item We integrate graph reconstruction, weighted graph contrastive learning, and semantic-structural alignment into MOTIF for robust sparse-supervision learning.
\item We conduct experiments under cold-start multimodal settings to validate MOTIF and its key components.
\end{itemize}
%%%%%%%%%%%%%%%%%%%%%%%%%%%%%%%%%%%%%%%%%%%%%%%%%%%%%%%%%%%%
\section{Related Work}

\subsection{Multimodal Recommendation}

Multimodal recommendation incorporates item content such as text, images, video frames, and attributes. MMGCN constructs modality-aware user-item graphs~\cite{wei2019mmgcn}, GRCN refines noisy feedback graphs~\cite{wei2020grcn}, LATTICE learns modality-aware item graphs~\cite{zhang2021lattice}, and FREEDOM improves robustness by freezing and denoising graph structures~\cite{zhou2023freedom}. Most methods, however, build item relations from feature similarity or latent correlations~\cite{miao2026universalrefinement}. In cold-start scenarios, similar appearance may hide different functions, causing semantic drift. MOTIF instead uses LLM-derived motivation semantics to capture transferable functional relations.

\subsection{Graph Collaborative Filtering and Graph Contrastive Learning}

Graph collaborative filtering learns representations over user-item graphs. NGCF and LRGCCF use graph neural or residual propagation~\cite{wang2019ngcf,chen2020lrgccf}, while LightGCN keeps only neighborhood propagation and layer-wise aggregation~\cite{he2020lightgcn}. Graph contrastive methods such as SGL~\cite{wu2021sgl}, NCL~\cite{lin2022ncl}, and SimGCL~\cite{yu2022simgcl} improve robustness through augmented or perturbed views. However, they usually assume reliable observed neighborhoods, which is weak for sparse users and cold items. WeightedGCL recalibrates final-layer perturbations with squeeze-and-excitation~\cite{chen2025weightedgcl}. MOTIF applies this idea on a knowledge-enhanced reconstructed graph for cold-start topology.

\subsection{Cold-start Recommendation}

Cold-start recommendation mainly relies on side information, representation generation, or rapid adaptation. DropoutNet encourages content-based inference by removing collaborative inputs~\cite{volkovs2017dropoutnet}; MeLU uses few-shot preference estimation~\cite{lee2019melu}; MetaHIN combines heterogeneous information networks with meta-learning~\cite{lu2020metahin}; and CVAR generates cold-item representations from side information~\cite{zhao2022cvar}. These methods rarely reason about latent motivations or reconstruct graph structures for cold-item message passing. MOTIF instead addresses cold-start recommendation from both semantic and topological perspectives.

\subsection{LLMs for Recommendation}

LLMs provide semantic understanding and reasoning for recommendation~\cite{miao2026r3rec}. P5 formulates recommendation as text-to-text prediction~\cite{geng2022p5}, and TALLRec adapts LLMs through instruction tuning~\cite{bao2023tallrec}. LLMRec augments profiles and graphs~\cite{wei2024llmrec}, RLMRec aligns LLM-derived semantics with collaborative embeddings~\cite{ren2024rlmrec}, and LMMRec extracts fine-grained motivations~\cite{di2026lmmrec}. The interaction among motivation reasoning, cold-start graph reconstruction, and graph contrastive learning remains underexplored. MOTIF converts LLM-derived motivations into item topology and distills them into graph embeddings without direct semantic injection.
%%%%%%%%%%%%%%%%%%%%%%%%%%%%%%%%%%%%%%%%%%%%%%%%%%%%%%%%%%%%
\section{Methodology}

\subsection{Problem Formulation}

Let $\mathcal{U}=\{u_1,u_2,\ldots,u_M\}$ and $\mathcal{I}=\{i_1,i_2,\ldots,i_N\}$ denote the user and item sets. The observed interaction matrix is
\begin{equation}
\mathbf{R}\in\{0,1\}^{M\times N},
\end{equation}
where $R_{ui}=1$ indicates that user $u$ has interacted with item $i$. Each item is associated with multimodal information
\begin{equation}
\mathcal{M}_i=\{t_i,v_i,a_i,\ldots\},
\end{equation}
where $t_i$, $v_i$, and $a_i$ denote textual, visual, and attribute-level information.

The goal of top-$K$ recommendation is to learn a scoring function
\begin{equation}
f:\mathcal{U}\times\mathcal{I}\rightarrow\mathbb{R},
\end{equation}
which ranks candidate items for each user. In cold-start scenarios, sparse users provide insufficient behavioral evidence, while cold items have limited collaborative connectivity. We formulate this task as a \emph{semantic-topological co-reconstruction} problem: semantic reconstruction infers user and item motivations from sparse multimodal evidence, and topological reconstruction converts such motivations into transferable item-item connectivity.This formulation extends conventional multimodal recommendation by treating multimodal content as motivation-level evidence rather than only auxiliary item features. The reconstructed topology is used as an explicit propagation structure, enabling cold items to receive collaborative signals through semantically and functionally related neighbors.

\subsection{Framework Overview}

\begin{figure}[t]
\centering
\IfFileExists{framework.pdf}{
\includegraphics[width=0.98\linewidth]{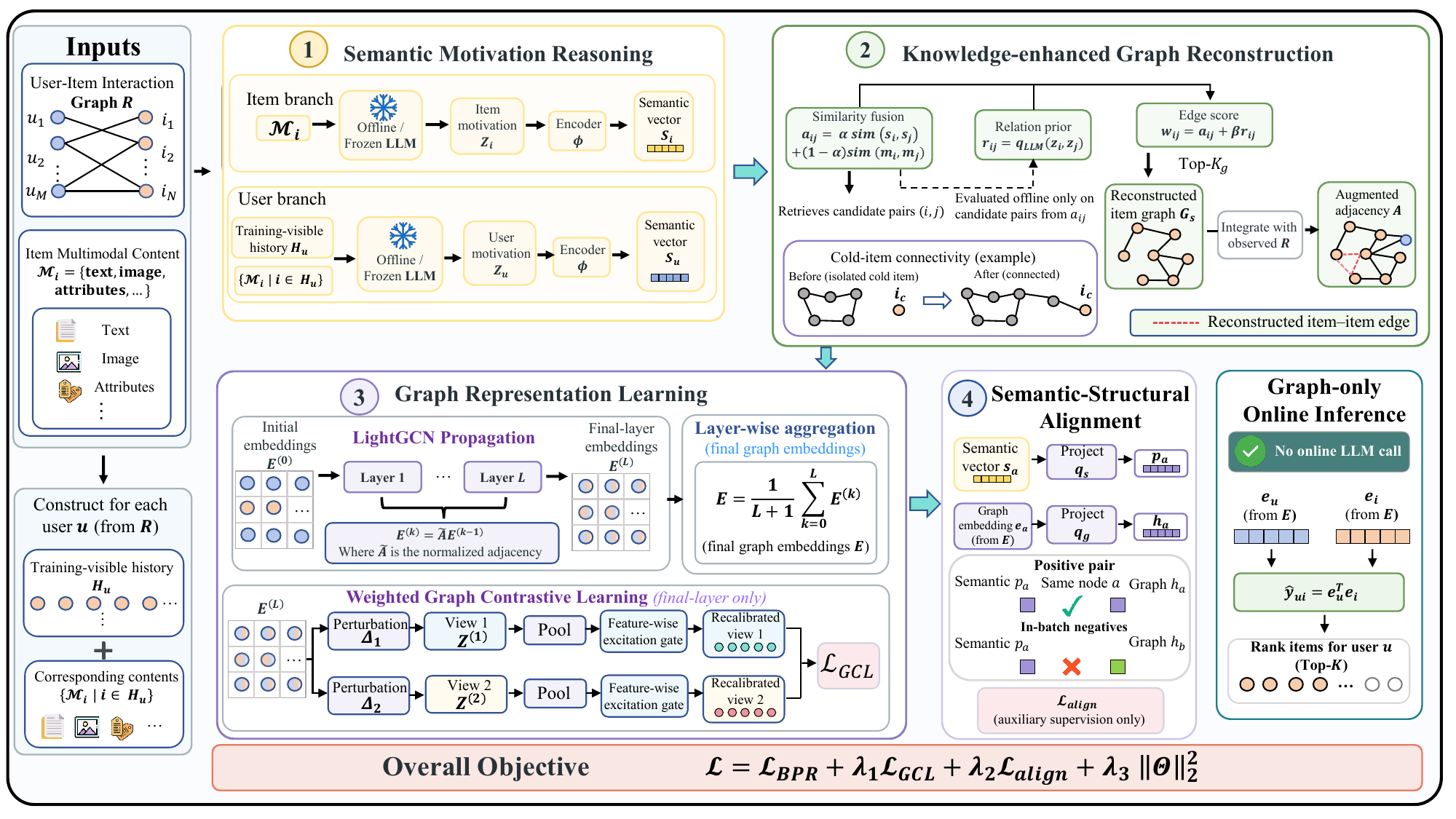}
}{
\fbox{\parbox{0.92\textwidth}{\centering Framework figure placeholder. Upload the framework PDF with the expected filename to replace this box.}}
}
\caption{Overall framework of the proposed semantic-topological co-reconstruction model.}
\label{fig_framework}
\end{figure}

Figure~\ref{fig_framework} illustrates MOTIF. It contains Semantic Motivation Reasoning, Knowledge-enhanced Graph Reconstruction, Weighted Graph Contrastive Learning, and Semantic-Structural Alignment. LLM-generated semantics guide graph reconstruction and alignment supervision, while final prediction uses only graph embeddings.

%%%%%%%%%%%%%%%%%%%%%%%%%%%%%%%%%%%%%%%%%%%%%%%%%%%%%%%%%%%%
\subsection{Semantic Motivation Reasoning}

Semantic Motivation Reasoning converts sparse multimodal evidence into compact motivation vectors. The LLM is used only offline, and its outputs are encoded before model training.

For item $i$, the LLM summarizes its functional role, usage scenario, target audience, and possible complementary or substitutable relations from $\mathcal{M}_i$:
\begin{equation}
z_i=g_{\mathrm{LLM}}(\mathcal{M}_i),
\end{equation}
where $z_i$ denotes the generated item motivation text. It is encoded as
\begin{equation}
\mathbf{s}_i=\phi(z_i), \quad \mathbf{s}_i\in\mathbb{R}^{d_s},
\end{equation}
where $\phi(\cdot)$ is a text encoder.

For user $u$, let
\begin{equation}
\mathcal{H}_u=\{i\mid R_{ui}=1\}
\end{equation}
denote the training-visible interaction history. To avoid information leakage, only training interactions are used. The user motivation text and semantic vector are obtained by
\begin{equation}
z_u=h_{\mathrm{LLM}}\left(\mathcal{H}_u,\{\mathcal{M}_i\mid i\in\mathcal{H}_u\}\right),
\end{equation}
\begin{equation}
\mathbf{s}_u=\phi(z_u), \quad \mathbf{s}_u\in\mathbb{R}^{d_s}.
\end{equation}

%%%%%%%%%%%%%%%%%%%%%%%%%%%%%%%%%%%%%%%%%%%%%%%%%%%%%%%%%%%%
\subsection{Knowledge-enhanced Graph Reconstruction}

To improve cold-item connectivity, we reconstruct a semantic-aware item graph by combining raw multimodal similarity, LLM-derived motivation semantics, and relation priors. Let $\mathbf{m}_i$ denote the original multimodal representation of item $i$. For item pair $(i,j)$, we compute
\begin{equation}
a_{ij}
=
\alpha \operatorname{sim}(\mathbf{s}_i,\mathbf{s}_j)
+
(1-\alpha)\operatorname{sim}(\mathbf{m}_i,\mathbf{m}_j),
\end{equation}
where $\operatorname{sim}(\cdot,\cdot)$ denotes cosine similarity, and $\alpha$ balances motivation semantics and raw multimodal similarity.

Since feature similarity may connect items with different functions, we introduce an offline relation prior:
\begin{equation}
r_{ij}=q_{\mathrm{LLM}}(z_i,z_j),
\end{equation}
where $q_{\mathrm{LLM}}(\cdot,\cdot)$ estimates functional relations such as complementarity, substitutability, or shared usage scenarios. For efficiency, $r_{ij}$ is computed only within a candidate set retrieved by $a_{ij}$.

The final edge score is
\begin{equation}
w_{ij}=a_{ij}+\beta r_{ij},
\end{equation}
and each item retains its top-$K_g$ neighbors:
\begin{equation}
\mathcal{N}_i^g=\operatorname{TopK}_{K_g}\left(\{w_{ij}\mid j\in\mathcal{I}, j\neq i\}\right).
\end{equation}
The reconstructed item graph is
\begin{equation}
\mathcal{G}_s=(\mathcal{I},\mathcal{E}_s),
\quad
\mathcal{E}_s=\{(i,j)\mid j\in\mathcal{N}_i^g\},
\end{equation}
with weighted adjacency matrix $\mathbf{A}_s$.

We integrate the reconstructed item graph with the original interaction graph:
\begin{equation}
\mathbf{A}
=
\begin{bmatrix}
\mathbf{0} & \mathbf{R}\\
\mathbf{R}^{\top} & \mathbf{A}_s
\end{bmatrix}.
\end{equation}
Let $\tilde{\mathbf{A}}=\mathbf{D}^{-\frac{1}{2}}\mathbf{A}\mathbf{D}^{-\frac{1}{2}}$ be the normalized adjacency matrix. We adopt LightGCN-style propagation:
\begin{equation}
\mathbf{E}^{(k)}=\tilde{\mathbf{A}}\mathbf{E}^{(k-1)}, \quad k=1,\ldots,L,
\end{equation}
and aggregate layer-wise embeddings:
\begin{equation}
\mathbf{E}=\frac{1}{L+1}\sum_{k=0}^{L}\mathbf{E}^{(k)}.
\end{equation}
The final user and item embeddings are denoted as $\mathbf{e}_u$ and $\mathbf{e}_i$.

%%%%%%%%%%%%%%%%%%%%%%%%%%%%%%%%%%%%%%%%%%%%%%%%%%%%%%%%%%%%
\subsection{Weighted Graph Contrastive Learning}

Weighted Graph Contrastive Learning improves robustness against uncertain reconstructed edges.

Instead of perturbing all graph layers, we perturb only the final-layer representations:
\begin{equation}
\mathbf{E}_1^{(L)}=\mathbf{E}^{(L)}+\Delta_1,
\quad
\mathbf{E}_2^{(L)}=\mathbf{E}^{(L)}+\Delta_2,
\end{equation}
where $\Delta_1$ and $\Delta_2$ are stochastic perturbations. The two contrastive views are
\begin{equation}
\mathbf{Z}^{(r)}
=
\frac{1}{L+1}
\left(
\sum_{k=0}^{L-1}\mathbf{E}^{(k)}
+
\mathbf{E}_r^{(L)}
\right),
\quad r\in\{1,2\}.
\end{equation}

For node $a$ in view $r$, let $\mathbf{z}_a^{(r)}\in\mathbb{R}^{d}$ denote its representation. We compute a compact feature statistic:
\begin{equation}
c_a^{(r)}
=
\operatorname{Pool}(\mathbf{z}_a^{(r)})
=
\frac{1}{d}\sum_{\ell=1}^{d} z_{a,\ell}^{(r)}.
\end{equation}
The excitation network produces a dimension-wise gate:
\begin{equation}
\mathbf{g}_a^{(r)}
=
\sigma\left(
\mathbf{W}_2
\delta\left(
\mathbf{W}_1 c_a^{(r)}
\right)
\right),
\quad
\mathbf{g}_a^{(r)}\in\mathbb{R}^{d}.
\end{equation}
The recalibrated representation is
\begin{equation}
\hat{\mathbf{z}}_a^{(r)}
=
\mathbf{z}_a^{(r)}
\odot
\mathbf{g}_a^{(r)}.
\end{equation}

For node $a$, the graph contrastive loss is
\begin{equation}
\mathcal{L}_{\mathrm{GCL}}
=
-
\sum_{a\in\mathcal{B}}
\log
\frac{
\exp(\operatorname{sim}(\hat{\mathbf{z}}_a^{(1)},\hat{\mathbf{z}}_a^{(2)})/\tau)
}{
\sum_{b\in\mathcal{B}}
\exp(\operatorname{sim}(\hat{\mathbf{z}}_a^{(1)},\hat{\mathbf{z}}_b^{(2)})/\tau)
},
\end{equation}
where $\mathcal{B}$ is a mini-batch of nodes and $\tau$ is the contrastive temperature.

%%%%%%%%%%%%%%%%%%%%%%%%%%%%%%%%%%%%%%%%%%%%%%%%%%%%%%%%%%%%
\subsection{Semantic-Structural Alignment}

LLM-derived semantic vectors and graph embeddings come from different representation spaces. To avoid direct fusion noise, we use an auxiliary alignment objective. For node $a$, let $\mathbf{s}_a$ and $\mathbf{e}_a$ denote its semantic vector and graph embedding. They are projected into a shared space:
\begin{equation}
\mathbf{p}_a=q_s(\mathbf{s}_a),
\quad
\mathbf{h}_a=q_g(\mathbf{e}_a),
\end{equation}
where $q_s(\cdot)$ and $q_g(\cdot)$ are learnable projection functions. The alignment loss is
\begin{equation}
\mathcal{L}_{\mathrm{align}}
=
-
\sum_{a\in\mathcal{B}}
\log
\frac{
\exp(\operatorname{sim}(\mathbf{p}_a,\mathbf{h}_a)/\tau_s)
}{
\sum_{b\in\mathcal{B}}
\exp(\operatorname{sim}(\mathbf{p}_a,\mathbf{h}_b)/\tau_s)
},
\end{equation}
where $\tau_s$ is the alignment temperature.

%%%%%%%%%%%%%%%%%%%%%%%%%%%%%%%%%%%%%%%%%%%%%%%%%%%%%%%%%%%%
\subsection{Prediction and Optimization}

The final recommendation score is computed only from graph embeddings:
\begin{equation}
\hat{y}_{ui}=\mathbf{e}_u^\top\mathbf{e}_i.
\end{equation}
Given a training triplet $(u,i^+,i^-)$, where $i^+$ is an observed item and $i^-$ is a sampled negative item, we adopt the Bayesian Personalized Ranking loss~\cite{rendle2009bpr}:
\begin{equation}
\mathcal{L}_{\mathrm{BPR}}
=
-
\sum_{(u,i^+,i^-)\in\mathcal{D}}
\log \sigma
\left(
\hat{y}_{ui^+}-\hat{y}_{ui^-}
\right).
\end{equation}

The overall objective is
\begin{equation}
\mathcal{L}
=
\mathcal{L}_{\mathrm{BPR}}
+
\lambda_1\mathcal{L}_{\mathrm{GCL}}
+
\lambda_2\mathcal{L}_{\mathrm{align}}
+
\lambda_3\|\Theta\|_2^2.
\end{equation}
Thus, prediction remains graph-based, while LLM-derived semantics supervise reconstruction and alignment.
%%%%%%%%%%%%%%%%%%%%%%%%%%%%%%%%%%%%%%%%%%%%%%%%%%%%%%%%%%%%
\section{Experiments}

We evaluate MOTIF in terms of overall accuracy, cold-start robustness, component effectiveness, graph quality, hyperparameter sensitivity, and efficiency.

\subsection{Experimental Setup}

\paragraph{Datasets.}
We conduct experiments on Amazon-Baby, Amazon-Sports, and MicroLens-50K. Amazon-Baby and Amazon-Sports contain user-item interactions with textual and visual item information, while MicroLens-50K contains short-video interactions with textual and visual-frame information. Interactions are split into training, validation, and test sets. User-side and item-side cold-start settings are constructed from the training graph. User-level LLM reasoning uses only training-visible interactions, and item-level reasoning uses only item-side multimodal content to prevent validation and test leakage.

\begin{table}[!ht]
\centering
\small
\setlength{\tabcolsep}{6pt}
\renewcommand{\arraystretch}{0.98}
\caption[Dataset Statistics]{Dataset statistics after preprocessing.}
\label{tab_dataset}
\begin{tabular}{lrrrr}
\toprule
Dataset & \#Users & \#Items & \#Interactions & Sparsity \\
\midrule
Amazon-Baby   & 19\,445 & 7\,050  & 160\,792 & 99.88\% \\
Amazon-Sports & 35\,598 & 18\,357 & 296\,337 & 99.95\% \\
MicroLens-50K & 50\,000 & 19\,220 & 359\,708 & 99.96\% \\
\bottomrule
\end{tabular}
\end{table}

\paragraph{Cold-start Protocol.}
For user-side cold-start evaluation, users with 1--2, 3--5, and more than 5 visible training interactions are regarded as extreme-cold, cold, and warm users, respectively. For item-side evaluation, cold items are created by masking their training interactions while keeping multimodal content available. The masked interactions are used only for validation and testing.

\paragraph{Baselines.}
We compare with representative baselines from five groups. Graph collaborative filtering baselines include MF-BPR~\cite{rendle2009bpr}, NGCF~\cite{wang2019ngcf}, LightGCN~\cite{he2020lightgcn}, UltraGCN~\cite{mao2021ultragcn}, LayerGCN~\cite{zhou2023layergcn}, and FKAN-GCF~\cite{xu2025fourierkan}. Graph contrastive learning baselines include SGL~\cite{wu2021sgl}, NCL~\cite{lin2022ncl}, SimGCL~\cite{yu2022simgcl}, LightGCL~\cite{cai2023lightgcl}, DCCF~\cite{ren2023dccf}, RecDCL~\cite{zhang2024recdcl}, BIGCF~\cite{zhang2024bigcf}, WeightedGCL~\cite{chen2025weightedgcl}, and LightCSCF~\cite{kai2026lightcscf}. Multimodal recommendation baselines include MMGCN~\cite{wei2019mmgcn}, GRCN~\cite{wei2020grcn}, LATTICE~\cite{zhang2021lattice}, FREEDOM~\cite{zhou2023freedom}, and CRANE~\cite{dai2026crane}. Cold-start baselines include DropoutNet~\cite{volkovs2017dropoutnet} and CVAR~\cite{zhao2022cvar}. LLM-enhanced baselines include LLMRec~\cite{wei2024llmrec}, RLMRec~\cite{ren2024rlmrec}, LMMRec~\cite{di2026lmmrec}, and RecGOAT~\cite{li2026recgoat}.

\paragraph{Evaluation Metrics.}
We adopt Recall@$K$ and NDCG@$K$ with $K\in\{10,20\}$. For each user, all items not observed in the training set are treated as candidate items. Each experiment is repeated with five random seeds, and the final results are reported as average values. Statistical significance is evaluated by paired $t$-test.

\paragraph{Implementation Details.}
All methods are implemented in PyTorch. The embedding dimension is set to 64. The number of graph propagation layers is tuned in $\{1,2,3,4\}$, the learning rate in $\{10^{-4},5\times10^{-4},10^{-3},5\times10^{-3}\}$, and the batch size in $\{1024,2048,4096\}$. For our method, LLM reasoning, semantic encoding, and graph reconstruction are performed offline, and no LLM call is used during online inference.

\subsection{Overall Performance}

Tables~\ref{tab_overall_graph} and~\ref{tab_overall_multimodal} report overall performance against graph-based, multimodal, cold-start, and LLM-enhanced baselines.

\begin{table}[!ht]
\centering
\scriptsize
\setlength{\tabcolsep}{4.0pt}
\renewcommand{\arraystretch}{1.00}
\caption[Graph Baselines]{Overall performance compared with graph-based baselines. Results are averaged over five random seeds; standard deviations are omitted for readability.}
\label{tab_overall_graph}
\begin{tabular*}{\textwidth}{@{\extracolsep{\fill}}lcccccc@{}}
\toprule
Method
& \multicolumn{2}{c}{Amz-Baby}
& \multicolumn{2}{c}{Amz-Sports}
& \multicolumn{2}{c}{MicroLens} \\
\cmidrule(lr){2-3}\cmidrule(lr){4-5}\cmidrule(lr){6-7}
& R@20 & N@20 & R@20 & N@20 & R@20 & N@20 \\
\midrule
\grouprow{Graph Collaborative Filtering}
MF-BPR    & 0.0487 & 0.0210 & 0.0533 & 0.0241 & 0.0460 & 0.0207 \\
NGCF      & 0.0591 & 0.0258 & 0.0650 & 0.0287 & 0.0552 & 0.0249 \\
LightGCN  & 0.0698 & 0.0301 & 0.0782 & 0.0344 & 0.0641 & 0.0290 \\
UltraGCN  & 0.0734 & 0.0322 & 0.0815 & 0.0363 & 0.0673 & 0.0307 \\
LayerGCN  & 0.0716 & 0.0314 & 0.0828 & 0.0359 & 0.0688 & 0.0314 \\
FKAN-GCF  & 0.0759 & 0.0331 & 0.0846 & 0.0372 & 0.0700 & 0.0319 \\
\addlinespace[0.6pt]
\grouprow{Graph Contrastive Learning}
SGL         & 0.0741 & 0.0325 & 0.0837 & 0.0369 & 0.0709 & 0.0323 \\
NCL         & 0.0768 & 0.0338 & 0.0861 & 0.0381 & 0.0724 & 0.0330 \\
SimGCL      & 0.0796 & 0.0349 & 0.0892 & 0.0397 & 0.0743 & 0.0338 \\
LightGCL    & 0.0785 & 0.0344 & 0.0884 & 0.0390 & 0.0751 & 0.0342 \\
DCCF        & 0.0812 & 0.0357 & 0.0905 & 0.0403 & 0.0762 & 0.0347 \\
RecDCL      & 0.0827 & 0.0361 & 0.0921 & 0.0408 & 0.0769 & 0.0351 \\
BIGCF       & 0.0834 & 0.0366 & 0.0917 & 0.0412 & 0.0775 & 0.0354 \\
WeightedGCL & 0.0858 & 0.0377 & 0.0948 & 0.0421 & 0.0796 & 0.0363 \\
LightCSCF   & 0.0871 & 0.0382 & 0.0960 & 0.0427 & 0.0805 & 0.0368 \\
\midrule
\methodname
& \textbf{0.1083} & \textbf{0.0472}
& \textbf{0.1146} & \textbf{0.0508}
& \textbf{0.0938} & \textbf{0.0429} \\
\midrule
Improv.
& 5.45\% & 6.07\% & 5.23\% & 5.83\% & 4.92\% & 5.15\% \\
\bottomrule
\end{tabular*}
\end{table}

\vspace{-2mm}

\begin{table}[!ht]
\centering
\scriptsize
\setlength{\tabcolsep}{4.0pt}
\renewcommand{\arraystretch}{1.00}
\caption[Multimodal \& LLM Baselines]{Overall performance compared with multimodal, cold-start, LLM-enhanced, and recent baselines. Results are averaged over five random seeds; standard deviations are omitted for readability.}
\label{tab_overall_multimodal}
\begin{tabular*}{\textwidth}{@{\extracolsep{\fill}}lcccccc@{}}
\toprule
Method
& \multicolumn{2}{c}{Amz-Baby}
& \multicolumn{2}{c}{Amz-Sports}
& \multicolumn{2}{c}{MicroLens} \\
\cmidrule(lr){2-3}\cmidrule(lr){4-5}\cmidrule(lr){6-7}
& R@20 & N@20 & R@20 & N@20 & R@20 & N@20 \\
\midrule
\grouprow{Multimodal Recommendation}
MMGCN   & 0.0641 & 0.0277 & 0.0712 & 0.0314 & 0.0598 & 0.0271 \\
GRCN    & 0.0739 & 0.0320 & 0.0808 & 0.0357 & 0.0664 & 0.0303 \\
LATTICE & 0.0859 & 0.0375 & 0.0950 & 0.0424 & 0.0731 & 0.0334 \\
FREEDOM & 0.0976 & 0.0420 & 0.1034 & 0.0455 & 0.0784 & 0.0357 \\
CRANE   & 0.0993 & 0.0431 & 0.1055 & 0.0468 & 0.0810 & 0.0371 \\
\addlinespace[0.6pt]
\grouprow{Cold-start Recommendation}
DropoutNet & 0.0612 & 0.0264 & 0.0681 & 0.0298 & 0.0570 & 0.0258 \\
CVAR       & 0.0708 & 0.0309 & 0.0774 & 0.0341 & 0.0649 & 0.0294 \\
\addlinespace[0.6pt]
\grouprow{LLM-enhanced Recommendation}
LLMRec  & 0.0934 & 0.0408 & 0.1012 & 0.0447 & 0.0828 & 0.0376 \\
RLMRec  & 0.0961 & 0.0419 & 0.1039 & 0.0459 & 0.0850 & 0.0388 \\
LMMRec  & 0.1008 & 0.0438 & 0.1070 & 0.0473 & 0.0879 & 0.0401 \\
RecGOAT & 0.1027 & 0.0445 & 0.1089 & 0.0480 & 0.0894 & 0.0408 \\
\midrule
\methodname
& \textbf{0.1083} & \textbf{0.0472}
& \textbf{0.1146} & \textbf{0.0508}
& \textbf{0.0938} & \textbf{0.0429} \\
\midrule
Improv.
& 5.45\% & 6.07\% & 5.23\% & 5.83\% & 4.92\% & 5.15\% \\
\bottomrule
\end{tabular*}
\end{table}

\vspace{2pt}

MOTIF consistently outperforms all baseline groups, validating the benefit of reconstructing cold-start topology and aligning motivation semantics with collaborative graph embeddings.

\subsection{Cold-start Performance}

To evaluate cold-start robustness, Table~\ref{tab_cold} reports performance on extreme-cold users, cold users, and cold items. We compare with representative baselines from graph collaborative filtering, graph contrastive learning, multimodal recommendation, and LLM-enhanced recommendation.

\vspace{2pt}

\begin{table}[!ht]
\centering
\scriptsize
\setlength{\tabcolsep}{3.4pt}
\renewcommand{\arraystretch}{0.94}
\caption[Cold-start Performance]{Cold-start performance comparison in terms of Recall@20 and NDCG@20. Results are averaged over five random seeds; standard deviations are below $0.0016$ and omitted for readability.}
\label{tab_cold}
\begin{tabular*}{\textwidth}{@{\extracolsep{\fill}}lcccccc@{}}
\toprule
Method
& \multicolumn{2}{c}{Amz-Baby}
& \multicolumn{2}{c}{Amz-Sports}
& \multicolumn{2}{c}{MicroLens} \\
\cmidrule(lr){2-3}\cmidrule(lr){4-5}\cmidrule(lr){6-7}
& R@20 & N@20 & R@20 & N@20 & R@20 & N@20 \\
\midrule
\grouprow{Extreme-cold users}
LightGCN    & 0.0291 & 0.0118 & 0.0327 & 0.0138 & 0.0254 & 0.0106 \\
WeightedGCL & 0.0348 & 0.0146 & 0.0389 & 0.0167 & 0.0301 & 0.0129 \\
FREEDOM     & 0.0412 & 0.0173 & 0.0461 & 0.0198 & 0.0338 & 0.0145 \\
LLMRec      & 0.0467 & 0.0201 & 0.0518 & 0.0227 & 0.0402 & 0.0176 \\
LMMRec      & 0.0521 & 0.0224 & 0.0573 & 0.0251 & 0.0441 & 0.0193 \\
\methodname & \textbf{0.0628} & \textbf{0.0271} & \textbf{0.0689} & \textbf{0.0304} & \textbf{0.0529} & \textbf{0.0233} \\
Improv.     & 20.54\% & 20.98\% & 20.24\% & 21.12\% & 19.95\% & 20.73\% \\
\addlinespace[1pt]

\grouprow{Cold users}
LightGCN    & 0.0438 & 0.0184 & 0.0489 & 0.0211 & 0.0372 & 0.0162 \\
WeightedGCL & 0.0507 & 0.0216 & 0.0561 & 0.0244 & 0.0428 & 0.0188 \\
FREEDOM     & 0.0579 & 0.0247 & 0.0638 & 0.0279 & 0.0471 & 0.0207 \\
LLMRec      & 0.0612 & 0.0263 & 0.0681 & 0.0298 & 0.0524 & 0.0231 \\
LMMRec      & 0.0668 & 0.0287 & 0.0729 & 0.0321 & 0.0563 & 0.0249 \\
\methodname & \textbf{0.0741} & \textbf{0.0319} & \textbf{0.0808} & \textbf{0.0357} & \textbf{0.0627} & \textbf{0.0277} \\
Improv.     & 10.93\% & 11.15\% & 10.84\% & 11.21\% & 11.37\% & 11.25\% \\
\addlinespace[1pt]

\grouprow{Cold items}
LightGCN    & 0.0223 & 0.0091 & 0.0251 & 0.0104 & 0.0198 & 0.0082 \\
WeightedGCL & 0.0276 & 0.0114 & 0.0308 & 0.0131 & 0.0237 & 0.0099 \\
FREEDOM     & 0.0389 & 0.0164 & 0.0432 & 0.0186 & 0.0314 & 0.0134 \\
LLMRec      & 0.0431 & 0.0183 & 0.0479 & 0.0208 & 0.0367 & 0.0159 \\
LMMRec      & 0.0488 & 0.0209 & 0.0537 & 0.0234 & 0.0408 & 0.0178 \\
\methodname & \textbf{0.0613} & \textbf{0.0264} & \textbf{0.0671} & \textbf{0.0297} & \textbf{0.0512} & \textbf{0.0226} \\
Improv.     & 25.61\% & 26.32\% & 24.95\% & 26.92\% & 25.49\% & 26.97\% \\
\bottomrule
\end{tabular*}
\end{table}

MOTIF achieves the best performance across all cold-start settings, indicating that motivation reasoning benefits sparse users and reconstructed topology benefits cold items.

\subsection{Ablation Study}

We conduct ablation studies to verify the contribution of each core component. The variants include removing LLM-based reasoning (\textbf{w/o LLM-R}), removing knowledge-enhanced graph reconstruction (\textbf{w/o KGR}), replacing the reconstructed graph with a similarity-based graph (\textbf{Similarity-Graph}), removing weighted graph contrastive learning (\textbf{w/o WGCL}), removing feature-wise recalibration (\textbf{w/o SE}), removing semantic-structural alignment (\textbf{w/o Align}), and directly fusing LLM semantic embeddings with graph embeddings for prediction (\textbf{Direct-Fusion}).

\vspace{2pt}

\begin{table}[!ht]
\centering
\scriptsize
\setlength{\tabcolsep}{3.8pt}
\renewcommand{\arraystretch}{0.94}
\caption[Ablation Components]{Ablation study on key components. Results are averaged over five random seeds; per-cell standard deviations are below $0.0013$ and omitted for readability.}
\label{tab_ablation}
\begin{tabular*}{\textwidth}{@{\extracolsep{\fill}}lcccccc@{}}
\toprule
Variant
& \multicolumn{2}{c}{Amz-Baby}
& \multicolumn{2}{c}{Amz-Sports}
& \multicolumn{2}{c}{MicroLens} \\
\cmidrule(lr){2-3}\cmidrule(lr){4-5}\cmidrule(lr){6-7}
& R@20 & N@20 & R@20 & N@20 & R@20 & N@20 \\
\midrule
w/o LLM-R        & 0.1016 & 0.0443 & 0.1080 & 0.0479 & 0.0850 & 0.0389 \\
w/o KGR          & 0.1006 & 0.0438 & 0.1028 & 0.0456 & 0.0875 & 0.0400 \\
Similarity-Graph & 0.1030 & 0.0449 & 0.1099 & 0.0487 & 0.0902 & 0.0413 \\
w/o WGCL         & 0.1036 & 0.0452 & 0.1101 & 0.0488 & 0.0881 & 0.0403 \\
w/o SE           & 0.1068 & 0.0465 & 0.1133 & 0.0502 & 0.0920 & 0.0421 \\
w/o Align        & 0.1046 & 0.0456 & 0.1113 & 0.0493 & 0.0913 & 0.0417 \\
Direct-Fusion    & 0.1021 & 0.0445 & 0.1093 & 0.0485 & 0.0907 & 0.0415 \\
\methodname       & \textbf{0.1083} & \textbf{0.0472} & \textbf{0.1146} & \textbf{0.0508} & \textbf{0.0938} & \textbf{0.0429} \\
\bottomrule
\end{tabular*}
\end{table}

\vspace{2pt}

The ablation results confirm the contribution of each component. Removing LLM reasoning or graph reconstruction degrades performance, and the gap between \textbf{Similarity-Graph} and MOTIF shows the benefit of motivation-aware reconstruction. The decrease of \textbf{Direct-Fusion} supports using LLM semantics as auxiliary supervision rather than prediction input.

\subsection{Reconstructed Graph Analysis}

We compare the original interaction graph, a multimodal similarity graph, and the proposed reconstructed graph to examine the quality of the reconstructed topology.

\begin{table}[!htbp]
\centering
\small
\setlength{\tabcolsep}{5.0pt}
\renewcommand{\arraystretch}{1.00}
\caption[Graph Diagnosis]{Diagnostic analysis of reconstructed item graphs.}
\label{tab_graph_diagnostic}
\begin{tabular*}{0.82\textwidth}{@{\extracolsep{\fill}}lcccc@{}}
\toprule
Graph Type & Iso. Cold & Cold Deg. & NU@10 & Build Time \\
\midrule
Original Graph      & 0.412 & 0.7  & 0.083 & --      \\
Similarity Graph    & 0.061 & 14.3 & 0.214 & 86.4s   \\
Reconstructed Graph & 0.017 & 19.8 & 0.317 & 173.5s  \\
\bottomrule
\end{tabular*}
\vspace{-1mm}
\end{table}

Table~\ref{tab_graph_diagnostic} shows that the reconstructed graph reduces isolated cold items, increases their average degree, and achieves higher NU@10, indicating more useful cold-item topology than similarity-based graphs.

\FloatBarrier
\subsection{Hyperparameter Sensitivity}

We analyze five important hyperparameters: the graph contrastive loss weight $\lambda_1$, the semantic-structural alignment weight $\lambda_2$, the semantic-modality balance coefficient $\alpha$, the relation-prior weight $\beta$, and the number of reconstructed neighbors $K_g$. Each hyperparameter is varied while the others are fixed to their best validation values.

\begin{figure}[!h]
\centering
\vspace{-1mm}
\includegraphics[
width=0.98\linewidth
]{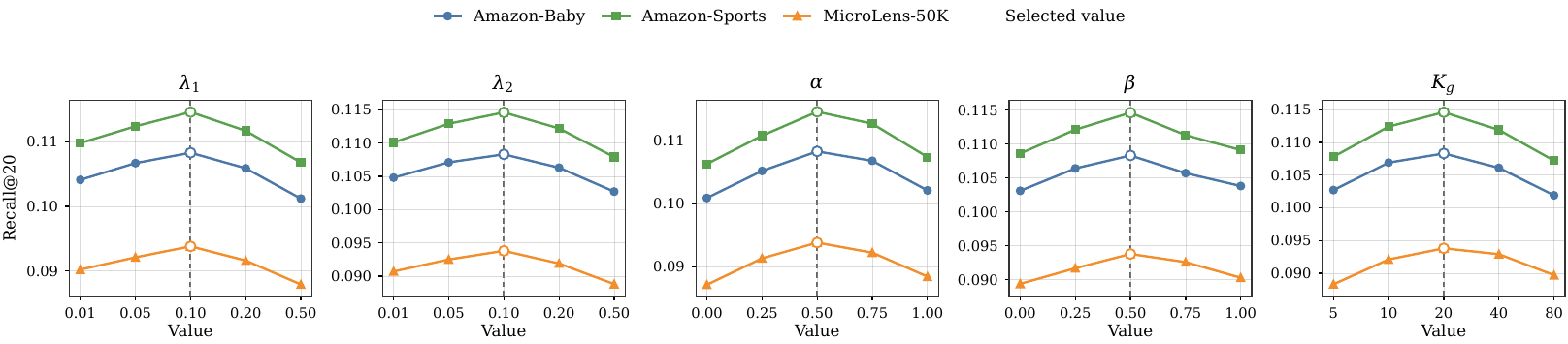}
\caption{Hyperparameter sensitivity analysis on Recall@20. Candidate indices correspond to increasing values of each hyperparameter, and the dashed vertical lines indicate the selected values.}
\label{fig_sensitivity}
\vspace{-2mm}
\end{figure}

Figure~\ref{fig_sensitivity} shows that moderate hyperparameter values generally perform best. Very small $\lambda_1$ or $\lambda_2$ weakens contrastive regularization or semantic alignment, whereas overly large values may dominate recommendation supervision. Intermediate $\alpha$, $\beta$, and $K_g$ better balance semantic connectivity and graph noise.

\FloatBarrier
\subsection{Efficiency Analysis}

We report offline preprocessing cost and online training/inference cost. Offline cost includes LLM reasoning, semantic encoding, and graph reconstruction, while online cost includes trainable parameters, training time, GPU memory, and inference latency. This separation is important because the expensive semantic reasoning stage is executed before training and does not participate in online recommendation.

\begin{table}[!htbp]
\centering
\scriptsize
\setlength{\tabcolsep}{3.8pt}
\renewcommand{\arraystretch}{0.92}
\caption[Efficiency Comparison]{Efficiency comparison. Offline LLM reasoning and graph reconstruction are performed before training and are not used during online inference.}
\label{tab_efficiency}
\begin{tabular*}{\textwidth}{@{\extracolsep{\fill}}lcccccc@{}}
\toprule
Method & Params & GPU Mem. & Train & Inference & Offline LLM & Recon. \\
\midrule
LightGCN    & 1.4M & 2.1GB & 11s & 0.8s & --    & --     \\
WeightedGCL & 1.6M & 2.4GB & 14s & 0.9s & --    & --     \\
FREEDOM     & 2.3M & 3.7GB & 23s & 1.1s & --    & 2.4min \\
LLMRec      & 2.1M & 3.2GB & 19s & 1.0s & 6.8h  & 3.1min \\
LMMRec      & 2.7M & 3.9GB & 27s & 1.2s & 9.4h  & 4.7min \\
RecGOAT     & 3.1M & 4.3GB & 31s & 1.3s & 11.2h & 5.9min \\
\methodname & 1.9M & 2.8GB & 17s & 0.9s & 8.1h  & 6.3min \\
\bottomrule
\end{tabular*}
\vspace{-1mm}
\end{table}

Table~\ref{tab_efficiency} shows that MOTIF introduces additional offline cost due to motivation reasoning and topology reconstruction, but these costs are amortized after preprocessing. Once semantic vectors and the reconstructed graph are cached, the model uses only graph embeddings for prediction, avoids online LLM calls, and keeps inference latency comparable to lightweight graph recommenders.
%%%%%%%%%%%%%%%%%%%%%%%%%%%%%%%%%%%%%%%%%%%%%%%%%%%%%%%%%%%%
\section{Conclusion}

This paper proposed MOTIF, an LLM-enhanced semantic-topological co-reconstruction framework for cold-start multimodal recommendation. MOTIF uses offline LLM reasoning to infer motivation semantics, reconstruct transferable item-item topology, and learn robust graph embeddings through weighted graph contrastive learning and semantic-structural alignment. Experiments on three datasets show consistent gains over graph-based, multimodal, cold-start, and LLM-enhanced baselines; ablations further validate motivation-aware reconstruction and auxiliary alignment. The results show that translating LLM-derived knowledge into graph structure and representation supervision outperforms direct semantic fusion. Prediction uses only graph embeddings, avoiding online LLM calls. Future work will explore adaptive graph priors and more efficient semantic reasoning.

%%%%%%%%%%%%%%%%%%%%%%%%%%%%%%%%%%%%%%%%%%%%%%%%%%%%%%%%%%%%

\begin{credits}
\subsubsection{\ackname}
This work was supported by the National College Student Innovation and Entrepreneurship Training Program under Project No.~202619145026.
\end{credits}

%%%%%%%%%%%%%%%%%%%%%%%%%%%%%%%%%%%%%%%%%%%%%%%%%%%%%%%%%%%%

\bibliographystyle{splncs04}
\bibliography{references}

\end{document}